\documentclass[preprint,prd,aps,showpacs,showkeys,nofootinbib]{revtex4-1}
\usepackage{amsmath}
\usepackage{amssymb}
\usepackage{amsfonts}
\usepackage{amssymb}
\usepackage{dcolumn}
\usepackage{bm}
\usepackage{graphicx}
\usepackage{xcolor}
\usepackage{array}
\usepackage{verbatim}
\usepackage{hyperref}

\begin{document}
	\title{ Generation of gravitational waves  in dynamical Chern-Simons gravity}
	\author{Zhi-Zhang Peng$^{1,2}$}
	\email{pengzhizhang@itp.ac.cn}
	
	\author{Zhen-Min Zeng$^{1,2}$}
	\email{cengzhenmin@itp.ac.cn}
	
	\author{Chengjie Fu$^{3}$}
	\email{fucj@ahnu.edu.cn}

	\author{Zong-Kuan Guo$^{1,2,4}$}
	\email{guozk@itp.ac.cn}
	
	\affiliation{$^1$CAS Key Laboratory of Theoretical Physics, Institute of Theoretical Physics, Chinese Academy of Sciences, P.O. Box 2735, Beijing 100190, China}
	
	\affiliation{$^2$School of Physical Sciences, University of Chinese Academy of Sciences, No.19A Yuquan Road, Beijing 100049, China}
	
	\affiliation{$^3$Department of Physics, Anhui Normal University, Wuhu, Anhui 241000, China}
	
	\affiliation{$^4$School of Fundamental Physics and Mathematical Sciences, Hangzhou Institute for Advanced Study, University of Chinese Academy of Sciences, Hangzhou 310024, China}
	
	\begin{abstract}
		We investigate gravitational waves (GWs) generated in a two-field inflationary model with a non-canonical kinetic term, in which the gravitational Chern-Simons term is coupled to a heavy dynamical field.
		In such a model, primordial GWs experience a period of resonant amplification for some modes.
		In addition, isocurvature perturbations suffer from a temporary tachyonic instability due to an effective negative mass,
		which source curvature perturbations, resulting in  large induced GWs. These two stochastic gravitationl wave backgrounds  correspond to different frequency bands, which are expected to be detected by future GW detectors such as SKA, LISA and Taiji.
	\end{abstract}
	\maketitle
	\section{introduction}
	Inflation, a compelling paradigm for the very early Universe, has successfully solved some theoretical  problems of the hot Big Bang model, such as the horizon and flatness problems~\cite{Guth:1980zm,Starobinsky:1980te,Linde:1981mu,Albrecht:1982wi}.
	Inflation predicts scalar perturbations which seed the large-scale structure formation of the Universe~\cite{Mukhanov:1981xt}
	and are supported by observations of the cosmic microwave background (CMB).
	On the other hand, tensor perturbations of the spacetime metric provide another crucial probe,
	i.e., primordial gravitational waves (GWs)~\cite{Starobinsky:1979ty,Rubakov:1982df}, which encode important information of the early Universe. Primordial GWs will induce the quadrupole anisotropies in the radiation field within the last scattering surface, causing B-mode polarization \cite{Verkhodanov:2021ddu}. Therefore, the precise measurement of B-mode
	polarization of CMB on large angular scales is considered to be the most promising method to detect primordial GWs. In addition, there are significant differences in the prediction of primordial GWs in assorted inflationary models, which is reflected in an important parameter tensor-to-scalar ratio $r$, directly related to the energy scale of inflation. The recent release of BICEP-Keck combined with the Planck 2018 result \cite{Planck:2018jri} gives an upper limit on the tensor-to-scalar ratio $r_{0.002}<0.035$ at a $95\%$ confidence level in the case of a scale-invariant power spectrum of tensor perturbations \cite{BICEP:2021xfz}. Consequently, primordial GWs are not only a smoking-gun probe of inflation, but also a powerful tool to distinguish various inflationary models.

	Primordial GWs, as a probe of gravitational symmetry, can also help us test the correctness of general relativity (GR). For a standard slow-roll inflation in
	the framework of GR, primordial GWs have two polarization modes which share exactly the same statistical properties and their power spectra coincide completely. However, the gravitational terms with parity violation are pervasive in abundant candidates of quantum gravity. One widely studied example is dynamical Chern-Simons (dCS) gravity \cite{Jackiw:2003pm,Alexander:2009tp}, in which a dynamical pseudoscalar field coupled to curvature via the Pontryagin density. This  modified gravity theory is motivated from the anomaly cancellation in heterotic string theory \cite{Green:1984sg} and from loop quantum gravity upon the promotion of the Barbero-Immirzi parameter to a field in the presence of matter \cite{Taveras:2008yf,Calcagni:2009xz}. As usual, the corrections on friction term  in Chern-Simons gravity induce the amplitude birefringence effect of GWs, associated with parity violation, which have been studied in previous works, eg. \cite{Lue:1998mq,Pogosian:2002dq,Balaji:2003sw,Alexander:2017jmt,Yunes:2010yf,Yagi:2017zhb,Alexander:2004wk,Fujita:2020iyx,Chu:2020iil,Yoshida:2017cjl,Li:2009rt,Cai:2016ihp,Odintsov:2022hxu,Zhao:2019xmm}. In the inflationary context, these works mainly focus on the influence of GWs on large scales, and only few works pay attention to possible small-scale observational effects.
	
	In Ref. \cite{Fu:2020tlw}, the authors successfully realize the resonant amplification of primordial GWs on small scales by introducing a periodic function of inflaton coupled to  Pontryagin density.
	However, it is not natural enough to put a sine function coupling by hand and lacks sufficient physical motivation.
	In the present paper, we investigate how to trigger the resonant amplification of GWs through natural oscillation of a scalar field during inflation in Chern-Simons gravity. We find the necessary conditions for this mechanism where the sub-dominated coupled scalar field, not inflaton, shall be so heavy that a large third derivative with respect to cosmic time during oscillation induces parametric resonance for some modes. Because we are interested in implementing this mechanism on small scales, we consider a non-canonical two-field inflationary model, in which the non-canonical term serves as a step-like function making the sub-dominated scalar field nearly frozen until the non-canonical effect vanishes. In addition, curvature perturbations are  also amplified in this model owing to the tachyonic instability of isocurvature perturbations, which has been discussed in previous papers \cite{Braglia:2020fms,Braglia:2020eai,Raveendran:2022dtb}. We also perform numerical calculation on curvature perturbations and find some new phenomena different from those in previous works, which will be discussed in detail later.
	To sum up, primordial GWs can be resonantly amplified in such a model considering the dCS coupling. Meanwhile, second order scalar-induced GWs will inevitably appear due to the enhancement of curvature perturbations.
	These two stochastic GW backgrounds (SGWBs) are located in different frequency bands and they are expected to be detected by future space-based GW detectors, such as LISA, Taiji and BBO.
	Joint observations of these two SGWBs provide a  breathtaking opportunity to test our model.

	The organization of the paper is as follows. In Sec.~\ref{II}, we present the amplification mechanism of primordial GWs in Chern-Simons gravity.
	In Sec.~\ref{III}, we introduce a feasible  two-field inflationary model and show numerical results about background and primordial GWs.
	In Sec.~\ref{IV}, we discuss curvature perturbations in this model.
	Sec.~\ref{V} is devoted to conclusion. Throughout the paper, we set $c=\hbar=1$, and the reduced Planck mass is defined as $M_{\rm p}=1/\sqrt{8 \pi G}$.

	\section{ gravitational waves in dynamical Chern-Simons gravity}
	\label{II}
	In this section, we discuss the possible significant observational effects of primordial gravitational waves in dCS gravity. We consider a scalar field $\chi$ coupled to a Chern-Simons term. The action for Chern-Simons gravity is as follows
	\begin{align}
		\label{action}
		S_{\rm {dCS}}=\frac
		{\alpha}{8} \int d^4 x\,\sqrt{-g}\chi R\tilde{R}\,,
	\end{align}
	where $\alpha$ is a dimensional quantity, and Pontryagin density $R\tilde{R}$ is defined by  $ \frac{1}{2} \epsilon^{\rho \sigma \alpha \beta} R^{\mu \nu}_{~~ \alpha \beta}R_{\nu \mu \rho \sigma} $ in which  $\epsilon^{\mu \nu \rho \sigma}$ is the four dimensional Levi--Civita tensor with $\epsilon^{0123}=-1/\sqrt{-g}$.

	In the flat Friedmann-Robertson-Walker (FRW) Universe in the presence of tensor perturbations, the perturbed metric reads
	\begin{align}
		ds^2 =-dt^2+a^2(t)(\delta_{ij}+h_{ij}(t,\mathbf{x}))dx^i dx^j\,,
	\end{align}
	with $h_{ij}$ representing the first-order transverse and traceless tensor perturbations, and $a(t)$,  the scale factor, denoting a function of cosmic time $t$. Proceeding to expand the action (\ref{action}) including well-known Einstein-Hilbert action up to second order in tensor perturbations, one can obtain
	\begin{align}
		\label{sh}
		S_h^{2}=\frac{M_{\rm p}^2}{8}\int dt d^3x a^3\big[\dot{h}_{ij}^2-\frac{1}{a^2}(\partial_k h_{ij})^2+\frac{\alpha\dot{\chi}}{aM^2_{\rm p}}\epsilon^{ijk}\dot{h}_{il}\partial_j\dot{h}_{kl}+\frac{\alpha\dot{\chi}}{a^3M^2_{\rm p}}\epsilon^{ijk}\partial^2h_{il}\partial_jh_{kl}\big]\,,
	\end{align}
	where $\epsilon^{ijk}$ is the Levi-Civita tensor. One could identify  the first two terms in the squared brackets in (\ref{sh}) coming from Einstein-Hilbert contribution while the remaining two terms represent the corrections brought by the Chern-Simons term. For the convenience of discussion, we expand $h_{ij}$ in terms of circular polarization basis in Fourier space
	\begin{align}
		h_{ij}(t,{\bf x}) = \sum_{A={\rm R, L}} \int \frac{d^3 {\bf k}}{(2\pi)^3} h_A(t, {\bf k})e^{i {\bf k}\cdot {\bf x}} e_{ij}^{A}({\bf k})\;,
	\end{align}
	where $e_{ij}^{A}$ is the circular polarization tensor, and $A=$L, R label the left-handed and right-handed polarization,respectively. The normalization and helicity conditions are
	\begin{subequations}
		\begin{align}
			e_{ij}^{A} (e_{ij}^B)^{\ast} &=2\delta^{AB},\\
			\epsilon_{i j n} k^j e_{im}^A &= i k\lambda_A e^A_{nm}\, \text{ with}\, \lambda_{\rm L}=-1,\, \lambda_{\rm R}=1.
		\end{align}
	\end{subequations}
	The quadratic action can now be rewritten as
	\begin{align}
		\label{sh1}
		S^2_h=\frac{M^2_{\rm p}}{8}\sum_{A={\rm R, L}} \int dt \int \frac{d^3{\bf k}}{(2\pi)^3}\left[1-\frac{\lambda_A k \alpha \dot{\chi}}{a M^2_{\rm p}}\right]\left[\dot{h}^2_A-\frac{k^2}{a^2}h^2_A\right]\,.
	\end{align}
	The action exists ghost modes if $\frac{\lambda_A k \alpha \dot{\chi}}{a M^2_{\rm p}}>1$.
	To avoid the appearance of vacuum instability, $D_A\equiv 1-\frac{\lambda_A k \alpha \dot{\chi}}{a M^2_{\rm p}}$ should be positive for both polarization modes~\cite{Dyda:2012rj}.
	
	Varying the action (\ref{sh1}) with respect to $h_{ij}$, we derive the equations of motion for $h_{ij}$,
	\begin{align}
		\ddot{h}_A+(3H+\frac{\dot{D}_A}{D_A})\dot{h}_A+\frac{k^2}{a^2}h_A=0\,.
		\label{GW_EOM_h}
	\end{align}
	Next, we theoretically analyze the necessary conditions to achieve effective amplification of $h_{ij}$ during inflation. We introduce a new variable $\bar{h}_A=a^{3/2}D_A^{1/2}h_A$, which yields a new expression
	\begin{align}\label{GW_EOM_X}
		\ddot {\bar{h}}_A + \left(\frac{k^2}{a^2} - \frac{\ddot F_A}{F_A}\right) \bar{h}_A=0\;,
	\end{align}
	where $F_A=a^{3/2}D_A^{1/2}$. As mentioned above, no ghost modes imply $|\frac{ k \alpha \dot{\chi}}{a M^2_{\rm p}}|<1$, thus, we can obtain
	\begin{align}\label{F}
		\frac{\ddot F_A}{F_A} \simeq  \frac{9}{4}H^2 +\frac{3}{2}\dot H -\frac{\rho_A}{2}\frac{k}{a} \left(\frac{1}{4}H^2\frac{\alpha\dot{\chi}}{aM^2_{\rm p}}+\frac{1}{2}\dot H\frac{\alpha\dot{\chi}}{aM^2_{\rm p}} + H\frac{\alpha\ddot{\chi}}{aM^2_{\rm p}} +\frac{\alpha\dddot{\chi}}{aM^2_{\rm p}}  \right)\;.
	\end{align}
	We consider a sub-dominated but heavy scalar field $\chi$, namely $m_\chi>H$, dramatically oscillates around the minimum of its effective potential during inflation, thus, $\chi$ obeys
	\begin{align}
		\ddot{\chi}+3H\dot{\chi}+m_\chi^2\chi=0\,.
	\end{align}
	The solution for the field $\chi$ asymptotically approaches the regime
	\begin{align}
		\chi(t)=X(t)\sin(m_\chi t).
	\end{align}
	Here $X(t)$ is the oscillation amplitude and we ignore the initial phase for simplicity. In the oscillating regime, $\ddot{\chi}$ changes rapidly since fast oscillation during a Hubble time, thus $\alpha\dddot{\chi}/aM^2_{\rm p} $ is dominated in  $\ddot{F}_A/F_A$ in Eq. (\ref{F}). Meanwhile, we roughly regard $X(t)$ as a constant in the derivative because of the slow change compared with the phase. So
	\begin{align}
		\frac{\ddot F_A}{F_A} \simeq \frac{\lambda_A}{2}\frac{k}{a}\frac{m_\chi^3 \alpha}{M^2_{\rm p}} X(t)\cos(m_\chi t)\,.
	\end{align}
	The equation of motion can be reduced to
	\begin{align}\label{h1}
		\ddot {\bar{h}}_A + \left(\frac{k^2}{a^2} - \frac{\lambda_A}{2}\frac{k}{a}\frac{m_\chi^3 \alpha}{M^2_{\rm p}} X(t)\cos(m_\chi t)\right) \bar{h}_A=0\;,
	\end{align}
	This equation describes a damping oscillator with a frequency $k_{phys}(\equiv k/a)$ driven by the polarization-dependent periodic force with the amplitude $\frac{\lambda_A}{2k_{phys}}\frac{m_\chi^3 \alpha}{M^2_{\rm p}} X(t)\cos(m_\chi t)$ and frequency $m_\chi$. We are concerned about the phenomenological significance of the above equation, such as whether efficient increase of $h_{ij}$ can be realized during oscillation to produce observable phenomena. Fortunately, the above equation can naturally lead to the well-known Mathieu equation, that is, we can expect the appearance of  resonant amplification for some modes. Making a change of the variable $m_\chi t=2z$ reduces Eq. (\ref{h1}) to celebrated Mathieu equation
	\begin{align}\label{mathieu}
		\frac{d^2\bar{h}_A}{dz^2} + \left[A_k -2q\cos(2z) \right] \bar{h}_A=0\;,
	\end{align}
	where
	\begin{align}
		A_k = \frac{k^2}{k_s^2a^2}\;, \qquad
		q = \frac{\lambda_Akm_\chi \alpha X}{aM_{\rm p}^2}=\frac{2k_sk}{aM^2} \;,
	\end{align}
	with $k_s=m_\chi/2$ and $M^2=\frac{M_{\rm p}^2}{\lambda_A \alpha X}$. According to the Floquet theory, $\bar{h}_A$ will present an exponential growth when the Floquet exponent $\mu_k$ has a real part, in which case the corresponding $(A_k,q)$ falls in the unstable band.
	In the present paper, we shall focus on narrow resonance where $q < 1$ and the most important instability band is located in the region around $|A_k-1| \sim \pm q$.
	
	In previous works \cite{Cai:2018tuh,Cai:2019bmk,Cai:2020ovp,Zhou:2020kkf,Peng:2021zon}, the control of parametric resonance, such as duration and strength of the resonances, were achieved by phenomenologically introducing an oscillatory term. For the current discussion, however, parametric resonance of tensor perturbations may arise from  oscillation of the scalar field during inflation, which has a more natural and sufficient physical motivation. During oscillation, one can estimate the amplitude amplification factor $\mathcal{A}(k)\approx \exp\left(\mu_k(t)k_s \Delta t\right)$, where the maximum of $\mu_k(t)$ is $q/2$ and the duration of resonance for a given mode within this band $\Delta t$ is about $\ln(\frac{1+q}{1-q})/2H$ estimated as $qH^{-1}$ in leading order in the case of narrow resonance. This leads to substantial growth of tensor perturbations $\bar{h}_A$ $\propto \exp(\frac{q^2m_\chi}{4H})$ when $q^2m_\chi \gg H $. Combined with $q<1$, usually, $q \sim O(0.1)$, only when $m_\chi$ is at least two orders of magnitude larger than $H$ the resonant amplification of $\bar{h}_A$ can be successfully realized.

	Such a heavy field is obviously not the inflaton because it cannot provide enough e-folds. In the two-field model, if $\chi$ is a sub-dominated spectator field, we  anticipate that such phenomena will occur on large scales which we are not interested in the present paper. To address this issue, a feasible model will be introduced in the next section, in which the non-canonical kinetic term is considered to control the time when the heavy field starts to oscillate.

	\section{Model and numerical results}
	\label{III}
	In this section, we discuss in detail a simple implementation of enhanced mechanism with a feasible two-field inflationary model. The action is given by	
	\begin{align}
		S[\phi,\chi]=\int d^4 x\,\sqrt{-g}\,\left[\frac{M_{\rm p}^2}{2} R-\frac{1}{2}(\partial \phi)^2-\frac{e^{2b(\phi)}}{2}(\partial\chi)^2-V(\phi,\chi)+\mathcal{L}_{\rm {{dCS}}} \right],
	\end{align}
	where $R$ is the Ricci scalar, $\phi$ is the inflaton field and $\chi$ is a sub-leading but heavy scalar field. $b(\phi)$ induces the interaction between two fields, and unambiguously $b(\phi)$ vanishes recovering standard kinetic terms.
	\begin{table}
		\begin{tabular}{>{\centering}p{2.cm}>{\centering}p{3cm}>{\centering}p{3cm}>{\centering}p{2cm}>{\centering}p{2cm}}
			\hline
			\hline
			$\textit{Set}$ & $\alpha/M^{-1}_{\rm p}$ & $\phi_c /M_{\rm p}$ & $b_1$& $\gamma/M_{\rm p}$ \tabularnewline
			\hline
			
			$1$ & $0.029$ & $4.94$ & $12$ & $10^{-2}$  \tabularnewline
			
			$2$ & $0.023$ & $4.65$ & $11$ & $10^{-2}$ \tabularnewline
			\hline
			\hline
		\end{tabular}
		\caption{Two parameter sets used in this paper.}
		\label{table:para}
	\end{table}
	This type of action  is motivated by generalized Einstein theories, which can also be naturally derived when $\chi$ is equivalent to an axionic component.
	
	We consider the following decoupled potential where inflation is mainly driven by the canonical field $\phi$ in the presence of a heavy non-canonical field $\chi$
	\begin{align}
		V(\phi,\chi)=U(\phi)+\frac{m^2_{\chi}}{2}\chi^2\,,
	\end{align}
	where $U(\phi)$ is in principle an arbitrary inflation potential favored by the current observational data.
	In our work the Starobinsky potential $U(\phi)=V_0[1-\exp(-\sqrt{2/3}\phi/M_{\rm p})]^2$ with $V_0=1.048\times 10^{-10}M^4_{\rm p}$ is considered as a concret example. In our model, $\chi$ is much heavier than the inflaton field $ \phi $, i.e., $m_{\chi} \gg m_{\phi}(\sim \sqrt{V_0}/M_{\rm p})$. We shall work with the following form for the function $b(\phi)$ \cite{Raveendran:2022dtb}
	\begin{align}
		b(\phi)=\frac{b_1}{2}\left\{1+\tanh[\frac{\phi-\phi_c}{\gamma}]\right\}\,,
	\end{align}
	where $b_1$ characterizes the strength of the coupling, $\phi_c$ is the turning point and $\gamma$ represents the speed of transformation.
	As discussed in the previous section, $b(\phi)$ is actually very similar to a step-like function, which makes the field $\chi$ ``light" at first and then ``heavy" at the turning point.

	The Friedmann equations for background and equations of motion in the FRW metric are
	\begin{align}\label{phi}		
		\ddot \phi + 3 H \dot \phi + V_\phi = b_\phi
		e^{2 b} \dot \chi^2 \,,
	\end{align}
	\begin{align}		
		\ddot \chi + (3 H + 2 b_\phi \dot \phi) \dot \chi
		+ e^{- 2 b} \, V_\chi = 0 \,,
	\end{align}
	\begin{align}\label{Hdot}
		\dot H = - \frac{1}{2 M_{\rm P}^2} \left[ \dot \phi^2 +
		e^{2 b} \dot \chi^2
		\right] \, .
	\end{align}
	\begin{figure*}
		\includegraphics[width=0.45\textwidth]{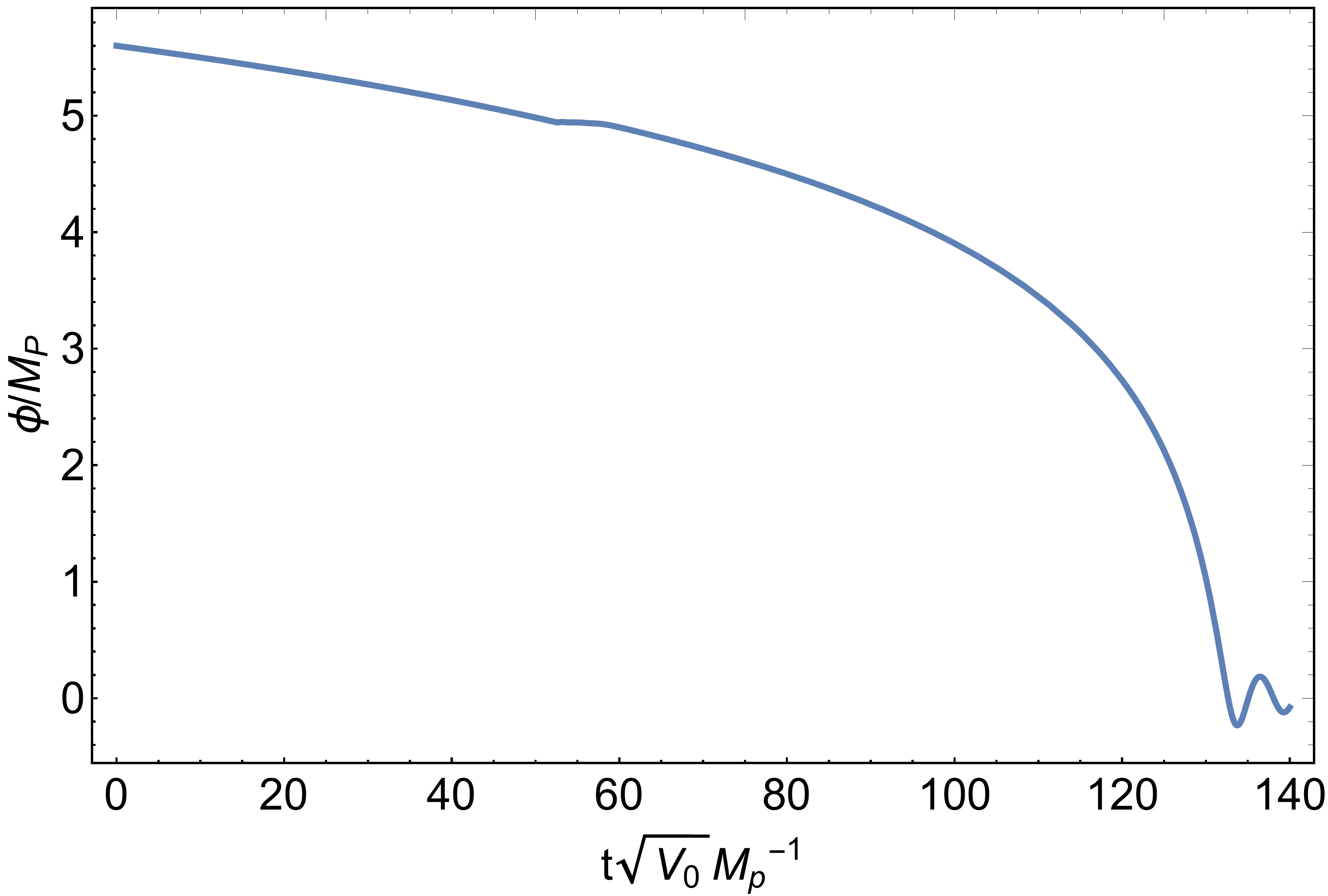}
		\includegraphics[width=0.5\textwidth]{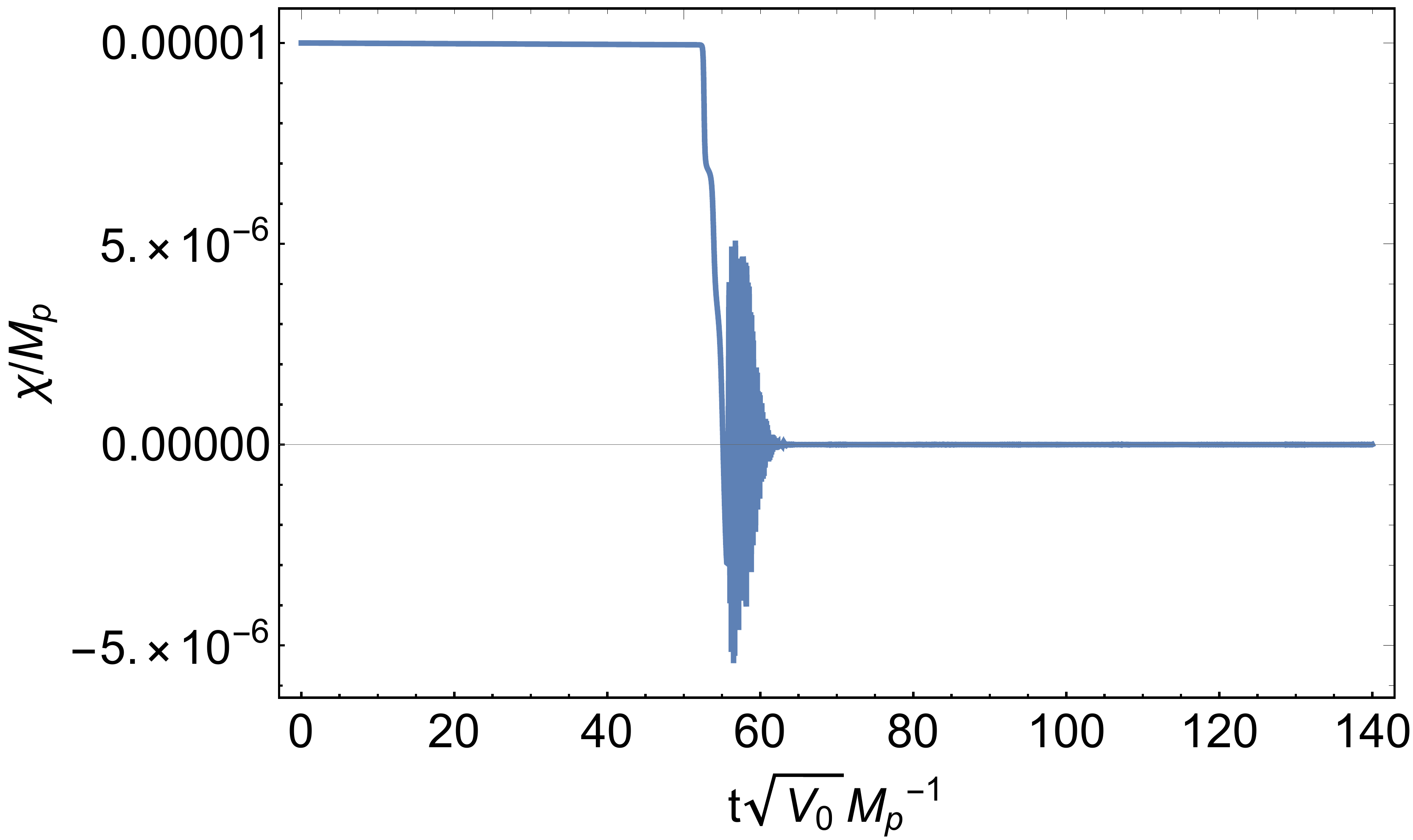}

		\caption{ Evolution of the inflaton field $\phi$ (left) and the scalar field $\chi$ (right) for Set 1. }
		\label{fig:eom}
	\end{figure*}

	Now we numerically calculate the evolution of background through Eqs.~\eqref{phi}-\eqref{Hdot}.
	In Fig. \ref{fig:eom}, we plot the evolution of $\phi$ and $\chi$
	for the illustrative case.
	Here $(m_{\chi} M_{\rm p})^2/V_0=4 \times 10^6$ and $V_0$ is fixed to produce the correct Planck normalization on CMB scales.
	We choose $\phi_i=5.6 M_\textup{p}$ and $\chi_i=10^{-5} M_\textup{p}$ for the initial values of the scalar fields. We discuss two sets of parameters that are relevant for observations in Table.~\ref{table:para}.

	As can be seen from Fig. \ref{fig:eom}, the lighter one of the two fields,
	i.e., the inflaton field $\phi$, rolls down its potential driving inflation  while the sub-dominated but heavier field $\chi$ remains nearly frozen at the first phase. The form of the function $b(\phi)$ induces a turning in the field space as the field $\phi$ approaches $\phi_c$. We find that the rolling of  $\chi$ from the turning point $\phi_c$ to the minimum is more complicated, which is of course because its sudden rolling takes away part of the kinetic energy of the inflaton.
	Nonetheless, it eventually settles into a stable oscillation.
	It is the rapid oscillation due to a large mass that induces parametric resonance of $h_{ij}$.
	
	\begin{figure*}
		\includegraphics[width=0.45\textwidth]{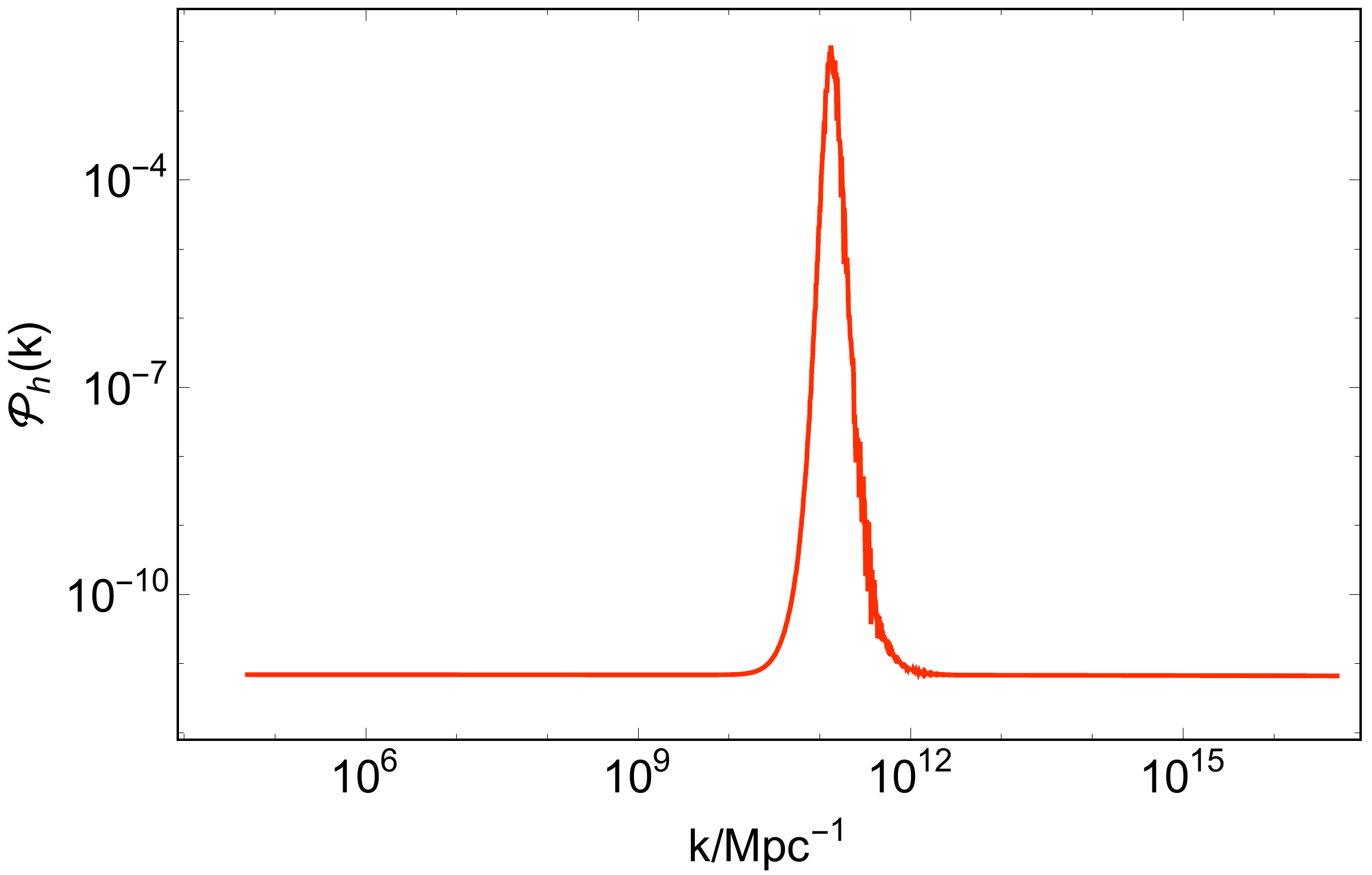}
		\includegraphics[width=0.45\textwidth]{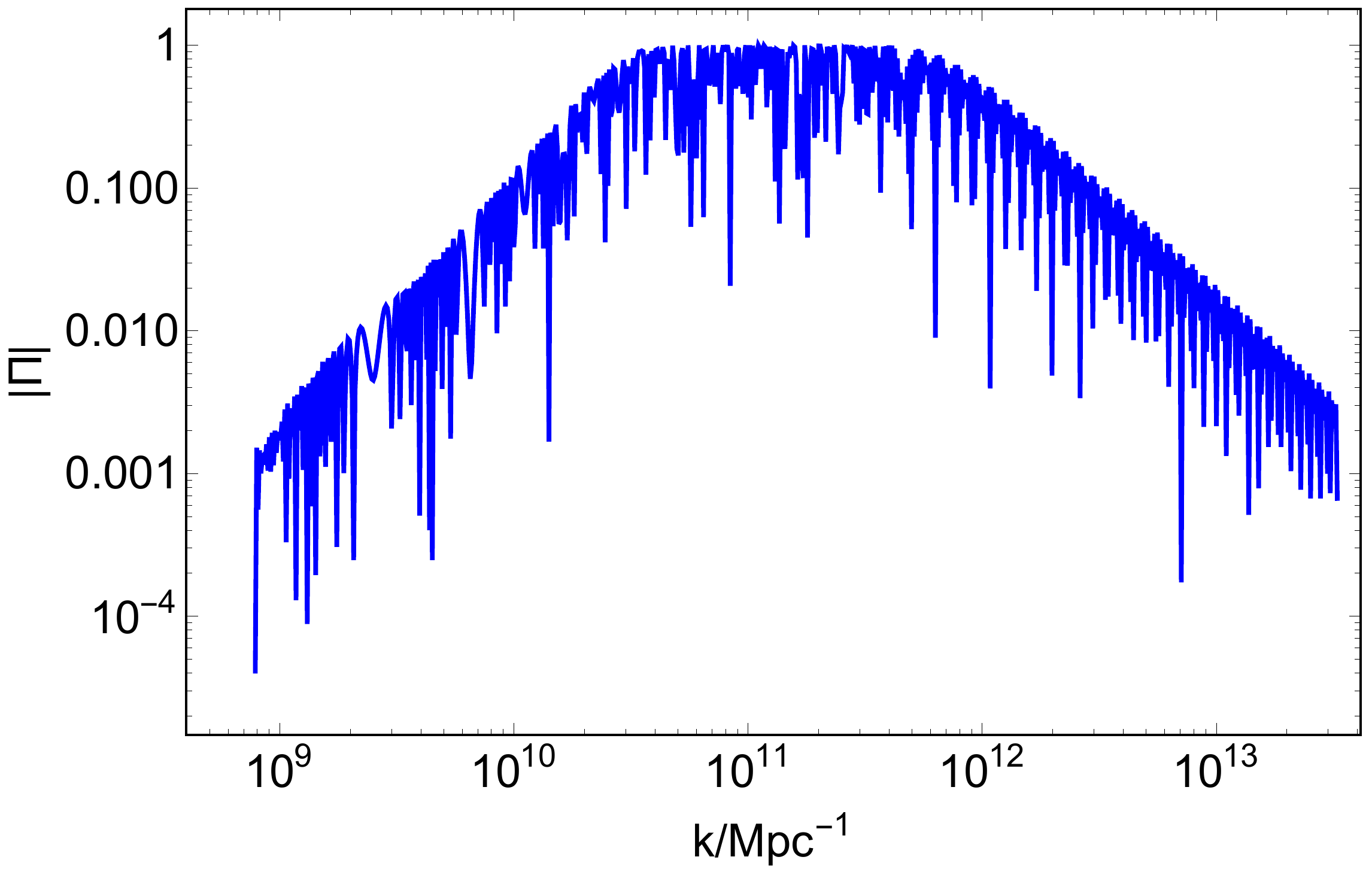}
		
		\caption{ Power spectrum of primordial GWs (left) and the absolute value of the degree of the circular polarization (right) for Set 1.}
		\label{fig:ph}
	\end{figure*}	
	
\begin{figure*}
		\includegraphics[width=0.8\textwidth]{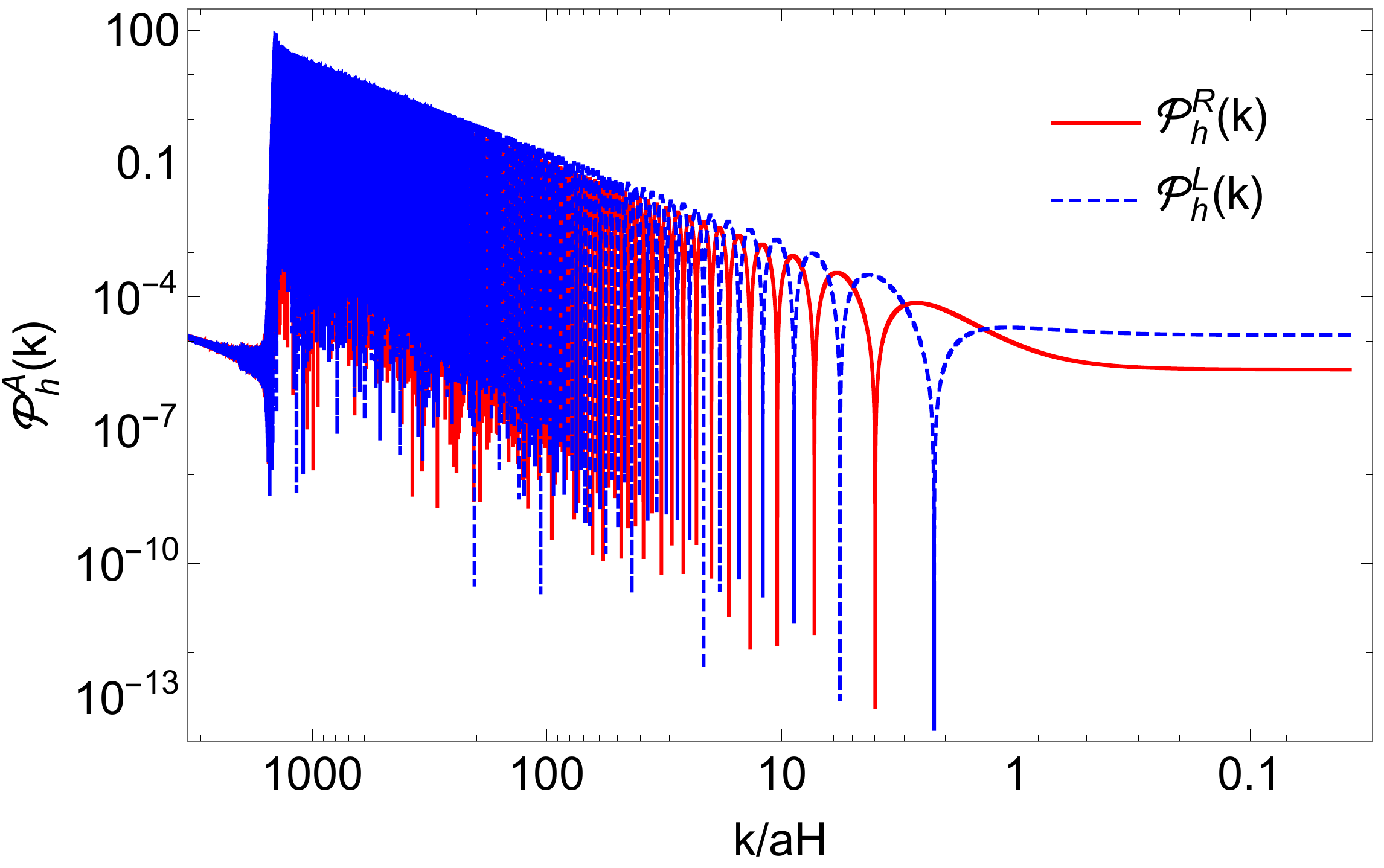}

		\caption{ Evolution of $\mathcal{P}^{\rm R}_h(k)$ and $\mathcal{P}^{\rm L}_h(k)$ for a resonant mode for Set 1.}
		\label{fig:ph_evol}
	\end{figure*}

	We normalize the scale factor $a(t)$ so that the pivot scale $k_*=0.05$ $\text{Mpc}^{-1}$ crosses the Hubble radius $N_*=60$ $e$-folds before the end of inflation. We depict the resulting power spectrum of tensor perturbations $\mathcal{P}_h=\sum_{A} \frac{k^3}{2\pi^2}|h^{A}_k|^2$  for Set 1 in Fig.~\ref{fig:ph}. Apparently, $\mathcal{P}_h$ shows a bump for at a certain mode. To characterize the chiral effect, we introduce the degree of the circular polarization, defined as
	\begin{align}
		\Pi=\frac{\mathcal{P}_h^\mathrm{R}-\mathcal{P}_h^\mathrm{L}}{\mathcal{P}_h^\mathrm{R}+\mathcal{P}_h^\mathrm{L}}\,.
	\end{align}
	The absolute value of the degree of the circular polarization is plotted in the right panel of Fig.~\ref{fig:ph}.
	By definition, $\Pi= 1,-1 $ are associated with  fully right- and left-handed polarized GWs, respectively.
	One can see that the $\Pi$ rapidly oscillates with $k$ within the resonant frequency band. This phenomenon originates from the  phase difference in the two resonant polarization modes after they cross the horizon, which stems directly from the difference in their equations of motion \eqref{GW_EOM_h} when the Chern-Simons coupling is considered (See Fig. \ref{fig:ph_evol}).
	
\begin{figure*}
		\includegraphics[width=0.8\textwidth]{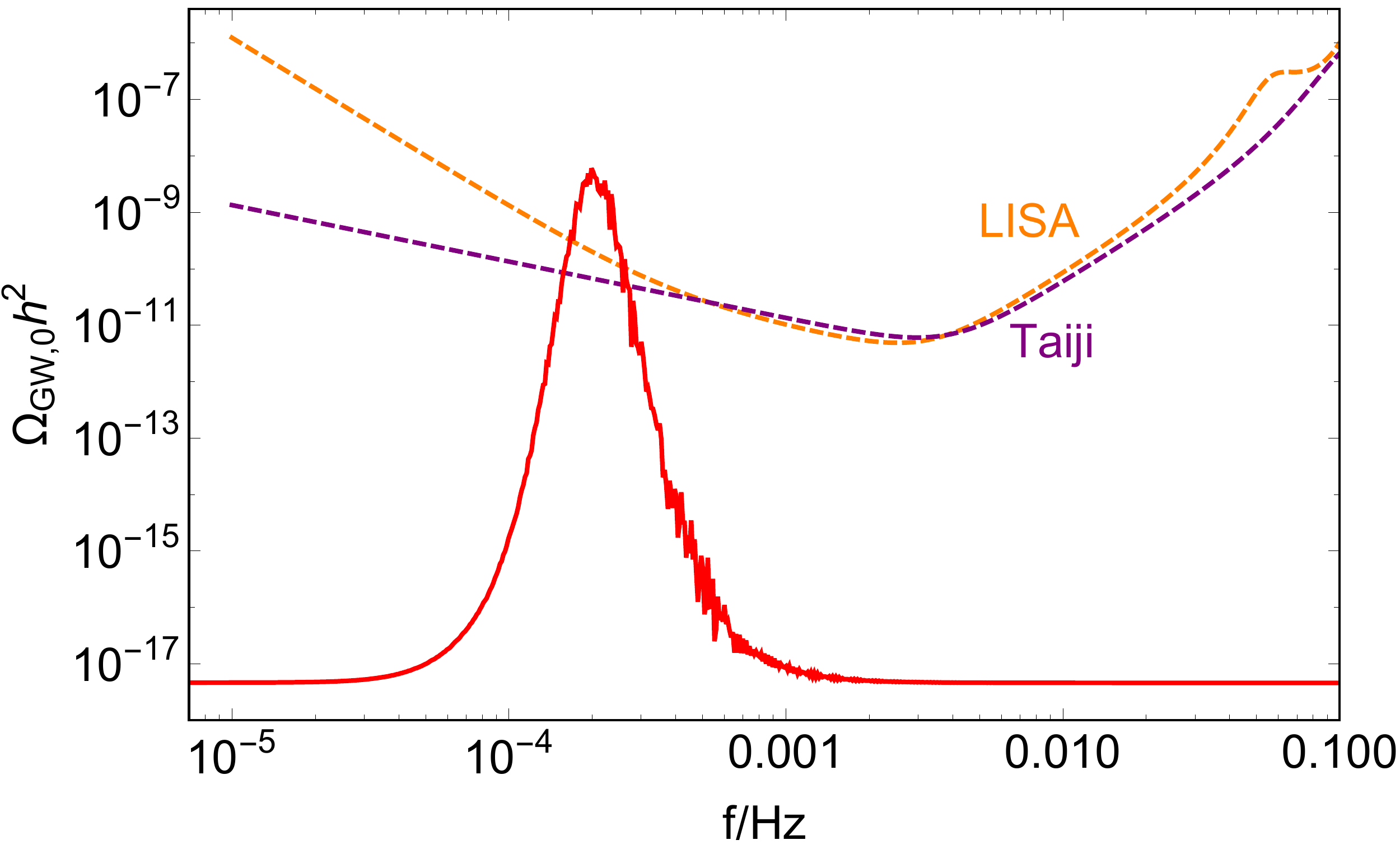}
		\caption{Energy spectrum of primordial GWs for Set 1.}
		\label{fig:omega_CSGW}
	\end{figure*}

	The present energy spectrum is related to the power spectrum  through \cite{Inomata:2021zel}
	\begin{align}
		\Omega_{\rm{GW,0}}(k)h^2=6.8\times10^{-7}\mathcal{P}_h(k)\,,
	\end{align}
	where $\mathcal{P}_h(k)$ is the total power spectrum of tensor perturbations.
	We show the resulting $\Omega_{\rm{GW,0}}h^2$ in Fig.~\ref{fig:omega_CSGW}.
	The peak of the energy spectrum exceeds the sensitivity curves of LISA~\cite{LISA:2017pwj} and Taiji~\cite{Ruan:2018tsw}.
	In~\cite{Seto:2020zxw,Orlando:2020oko} it has been pointed out that a parity violation signature with $|\Pi| \Omega_{\rm{GW}}(k)h^2 \sim 10^{-12} $ in an isotropic SGWB might be revealed by cross-correlating the data of LISA and Taiji.
	Accordingly, the chirality of primordial GWs predicted in the present paper are expected to be detected by the LISA-Taiji network in the future~\cite{Ruan:2019tje,Ruan:2020smc,Wang:2020dkc}.
	The SGWB with parity violation is unique, whose chirality can distinguish our model from other inflationary hypotheses.

	\section{Curvature perturbations and induced GWs}
	\label{IV}
	The two-field inflationary model in the present paper actually leads to increase of curvature perturbations sourcing an important production of GWs in the radiation-dominated era, which has been studied in~\cite{Raveendran:2022dtb}. However, due to the different parameter settings, the emergence of some new phenomena will be different from those in previous studies.
	In this section, we first elaborate on the basic formula for perturbations, and then present our results.

	\subsection{Curvature perturbations}
\begin{figure*}
		\includegraphics[width=0.8\textwidth]{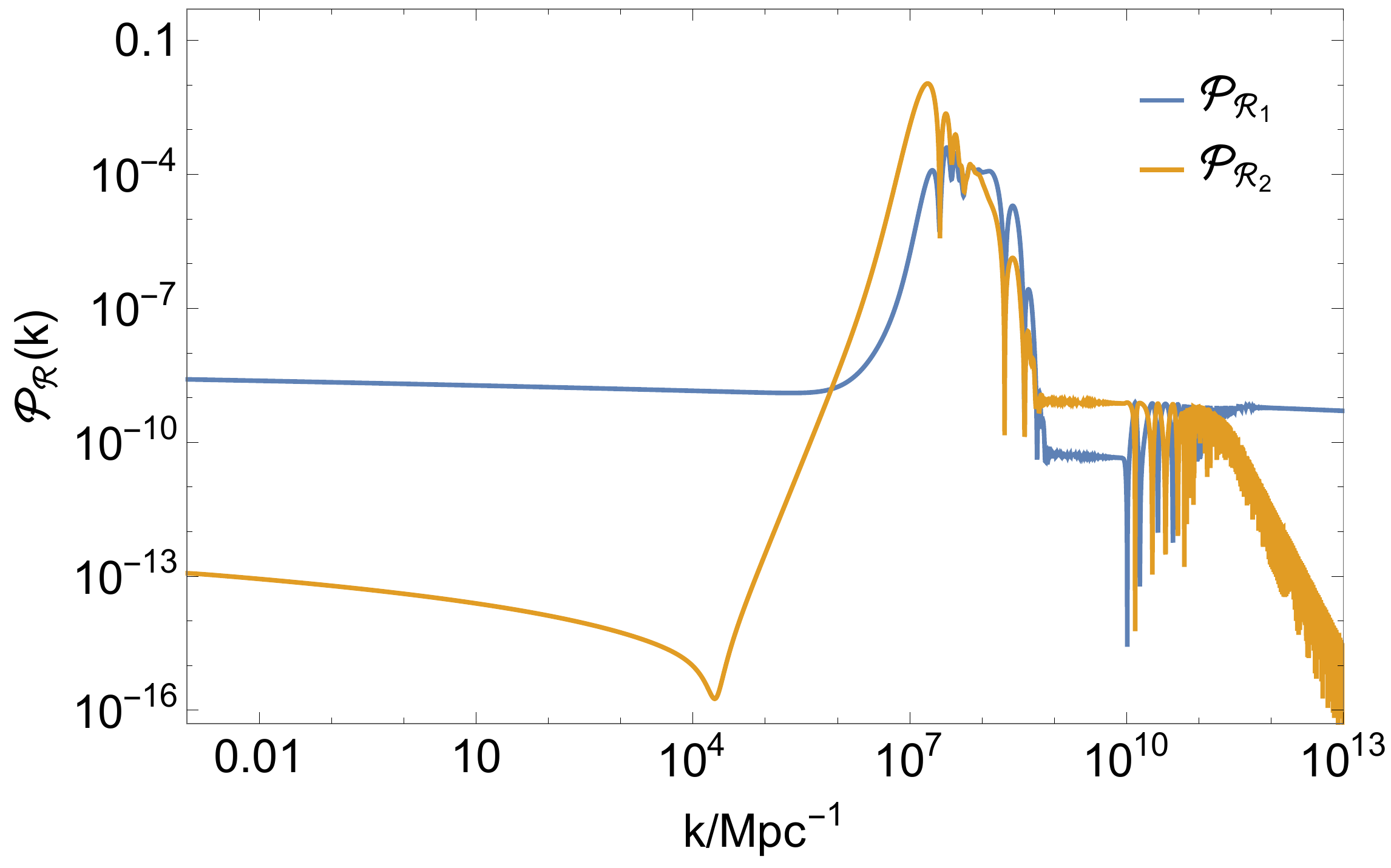}		
		\caption{ Power spectra of curvature perturbations corresponding to different initial conditions for Set 1.}
		\label{fig:pr12}		
	\end{figure*}

	We now discuss the linear perturbations in the Newtonian gauge. In the absence of anistropic
	stress, the perturbed FRW metric in the Newtonian gauge is of the form \cite{Mukhanov:1990me}
	\begin{align}
		\mathrm{d}s^2 = -( 1 + 2 \Phi) \mathrm{d}t^2 + a^2 (1 - 2 \Phi) \mathrm{d} {\bf x}^2 \, , \label{metric}
	\end{align}
	where $\Phi$ is the Newtonian potential characterizing  scalar perturbations.
	We decompose the scalar fields into their background parts and perturbations: $	\phi(t,{\bf x}) = \phi(t) +\delta\phi(t,{\bf x})$ and $	\chi(t,{\bf x}) = \chi(t) +\delta\chi(t,{\bf x})$. In order to interpret the evolution of cosmological perturbations conveniently, $\delta\phi$ and $\delta\chi$ are decomposed along the two directions respectively parallel and orthogonal to the homogeneous trajectory in the field space~\cite{Gordon:2000hv}.
	We call the projection parallel to the trajectory the adiabatic (or curvature) component, while the orthogonal one is equivalent with the entropy (or isocurvature) component. These quantities at the linear level are given by \cite{DiMarco:2002eb}
	\begin{subequations}
		\label{eq:ai}
		\begin{eqnarray}
			\label{adi}
			\delta\sigma&=&\cos\theta\, \delta\phi
			+\sin\theta\, {\rm e}^b\, \delta\chi,\\
			\label{iso}
			\delta s&=&-\sin\theta\,\delta\phi
			+\cos\theta\, {\rm e}^b\, \delta\chi,
		\end{eqnarray}
	\end{subequations}
	where $\cos\theta=\frac{\dot{\phi}}{\dot{\sigma}}$,
	$\sin\theta={\rm e}^b \frac{\dot{\chi}}{\dot{\sigma}}$ and
	$\dot{\sigma}^2=\dot{\phi}^2+{\rm e}^{2b}\dot{\chi}^2$.
	The following quantities are also defined for the convenience of discussion
	\begin{subequations}
		\begin{eqnarray}
			V_{\sigma\sigma} &=&V_{\phi\phi}\cos^2\theta
			+ {\rm e}^{-b} V_{\phi\chi} \sin 2 \theta
			+ {\rm e}^{-2 b} V_{\chi\chi} \sin^2\theta,\\
			V_{ss} &=&V_{\phi\phi}\sin^2\theta
			- {\rm e}^{-b} V_{\phi\chi}\sin 2 \theta
			+ {\rm e}^{-2 b} V_{\chi\chi}\cos^2\theta ,\\
			V_{\sigma s} &=&-V_{\phi\phi}\cos\theta\sin\theta
			+ {\rm e}^{-b} V_{\phi\chi} (\cos^2\theta-\sin^2\theta)
			+ {\rm e}^{-2 b} V_{\chi\chi}\cos\theta\sin\theta,
		\end{eqnarray}
	\end{subequations}
	with $	V_\sigma=V_\phi\cos\theta+{\rm e}^{-b}\,V_\chi\sin\theta$ and $	V_s=-V_\phi\sin\theta+{\rm e}^{-b}\,V_\chi\cos\theta$.
	Curvature perturbations and isocurvature perturbations are defined as
	\begin{align}
		\mathcal{R}\equiv\frac{H}{\dot{\sigma}}Q_\sigma \,,\quad \mathcal{S}\equiv\frac{H}{\dot{\sigma}}Q_ s\,.	
	\end{align}
	Here $Q_\sigma$ and $Q_s$ are the Mukhanov-Sasaki variables, which are given by $Q_{\sigma}\equiv \delta\sigma +  \frac{\dot\sigma}{H}\Phi$, and $Q_s=\delta s$. We note that $\delta s$ is automatically gauge-invariant from definition. The equations of motion for perturbations are given by \cite{Lalak:2007vi}
	\begin{eqnarray}
		\label{eom1d}
		&& \ddot{Q}_\sigma+3H\dot{Q}_\sigma
		+\left(\frac{k^2}{a^2}+M_{\sigma\sigma}\right)Q_\sigma +\frac{2V_s}{\dot\sigma}\dot{\delta s} + M_{\sigma s}\,\delta s
		=0\,,
		\\
		\label{eom2d}
		&& \ddot{\delta s}+3H\dot{\delta s}
		+\left(\frac{k^2}{a^2}+M_{ss}\right)\delta s -\frac{2V_s}{\dot\sigma}\dot{Q}_\sigma +M_{s\sigma} Q_\sigma=0 \, ,
	\end{eqnarray}
	with
	\begin{eqnarray}
		\label{cd01}
		M_{\sigma\sigma} &=& V_{\sigma\sigma}-\left(\frac{V_s}{\dot{\sigma}}\right)^2
		+2\frac{\dot{\sigma} V_\sigma}{M_P^2 H}
		+\frac{3\dot{\sigma}^2}{M_P^2}-\frac{\dot{\sigma}^4}{2M_P^4H^2}
		-b_\phi\left(s_\theta^2c_\theta V_\sigma+(c_\theta^2+1)s_\theta V_s\right)\,,
		\\
		M_{\sigma s} &=& 6H\frac{V_s}{\dot{\sigma}}+\frac{2V_\sigma V_s}{\dot{\sigma}^2}+2V_{\sigma s}+ \frac{\dot{\sigma} V_s}{M_P^2 H}
		+2b_\phi(s_\theta^3V_\sigma-c_\theta^3V_s)\,,
		\\
		M_{ss} &=& V_{ss} -\left(\frac{V_s}{\dot{\sigma}}\right)^2
		+b_\phi(1+s_\theta^2)c_\theta V_\sigma +b_\phi c_\theta^2s_\theta V_s - \dot{\sigma}^2(b_{\phi\phi}+b_\phi^2)\,,
		\\
		\label{cd04}
		M_{s\sigma} &=&
		-6H\frac{V_s}{\dot{\sigma}}-\frac{2V_\sigma V_s}{\dot{\sigma}^2}
		+\frac{\dot{\sigma} V_s}{M_P^2 H}\,,
	\end{eqnarray}
	where $s_\theta\equiv\sin\theta$ and $c_\theta\equiv\cos\theta$.		
	Generally, in order to ensure the statistical independence of $Q_\sigma$ and $\delta s$ at the deep horizon, we adopt the method of integrating twice to solve the above equations. We impose the Bunch-Davies initial conditions on $Q_\sigma$ and assume that the initial value of $\delta s$ is zero in first integration. Then we exchange the initial conditions and finally obtain two sets of solutions, $(\mathcal{R}_1,S_1)$ and $(\mathcal{R}_2,S_2)$.
	The curvature and isocuvature power spectra and cross power spectra are defined as
	\begin{subequations}
		\label{P1P2}
		\begin{eqnarray}
			\mathcal{P}_{\mathcal R}(k)
			&=&\frac{k^3}{2\pi^2}
			(\lvert\mathcal R_1\rvert^2+\lvert\mathcal R_2\rvert^2)
			=\mathcal{P}_{\mathcal R_1}(k)+\mathcal{P}_{\mathcal R_2}(k),
			\label{eq:PR}\\
			\mathcal{P}_{\mathcal S}(k)
			&=&\frac{k^3}{2\pi^2}
			(\lvert \mathcal S_1\rvert^2+\lvert \mathcal S_2\rvert^2),\\
			\mathcal{C}_{\mathcal {RS}}(k)
			&=&\frac{k^3}{2\pi^2}
			(\mathcal R^\ast_1\mathcal S_1+\mathcal R^\ast_2\mathcal S_2).
		\end{eqnarray}
	\end{subequations}
	\begin{figure*}
		\includegraphics[width=0.458\textwidth]{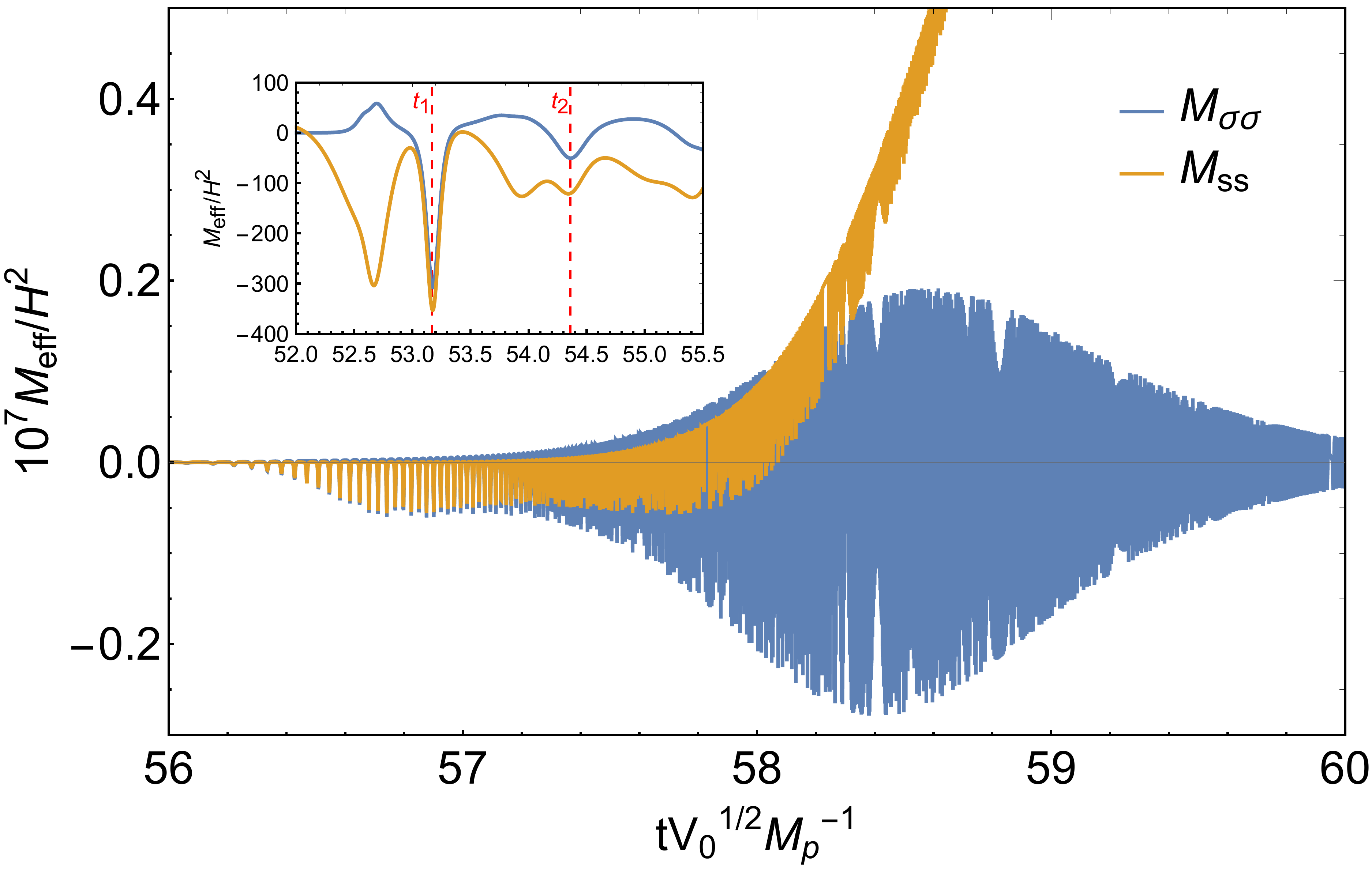}	
		\includegraphics[width=0.385\textwidth]{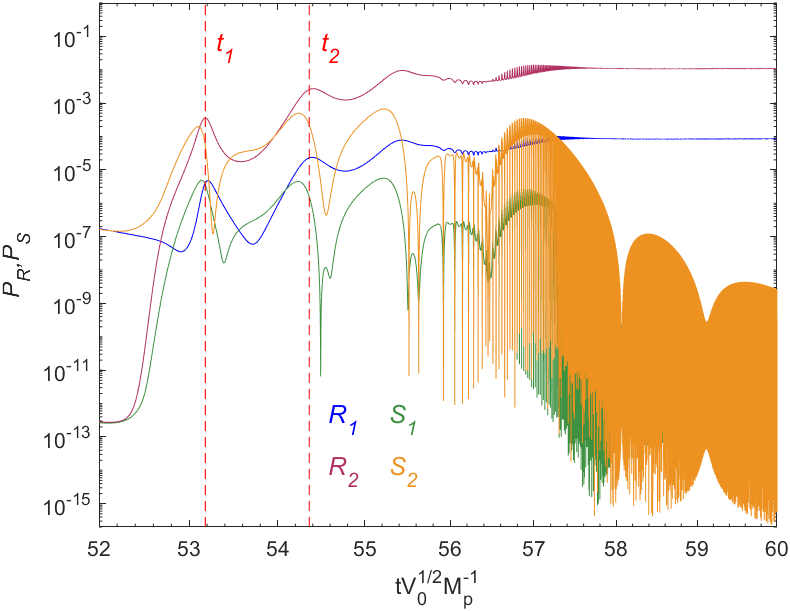}	
		\caption{ Evolution of the isocurvature mass and curvature mass (left) and the 		
			evolution of the power spectra $\mathcal{P}_{\mathcal{R}}=\frac{k^3}{2\pi^2}|\mathcal{R}|^2$ and $\mathcal{P}_{\mathcal{S}}=\frac{k^3}{2\pi^2}|\mathcal{S}|^2$  for the most enhanced mode $k_m$ (right) for Set 1.  }
		\label{fig:meff}		
	\end{figure*}
	The left panel of Fig.~\ref{fig:meff} plots the evolution of the isocurvature mass and curvature mass for Set 1.
	The corresponding  power spectra of curvature perturbations and isocurvature perturbations for the most enhanced mode $k_m$  are shown in the right panel of Fig. \ref{fig:meff}.
	We focus on the period when $\chi$ oscillates rapidly.
	We see that during a period of oscillation, the isocurvature mass $M_{ss}$ becomes temporarily negative triggering a tachyonic instability,
	which acts as a source leading to exponential growth in curvature perturbations.
	This fact is confirmed by our numerical computation showed in the right panel (see the evolution of $\mathcal{P}_{\mathcal{R}_2}$).
	Meanwhile, we note that around both $t_1$ and $t_2$ the curvature mass $M_{\sigma\sigma}$ also becomes negative
	which means that curvature perturbations suffer from the same tachyonic instability.
	A direct result of this interesting phenomenon is that curvature perturbations ($\mathcal{P}_{\mathcal{R}_1}$) experience exponential amplification without relying on isocurvature perturbations as well,
	which is a novel discovery different from that in the previous study.
	The efficiency of enhancement, however, is weaker than the former.
	As a result, the enhanced part of $\mathcal{P}_{\mathcal{R}_1}$ on small scales is washed out in the resulting power spectrum $\mathcal{P}_{\mathcal{R}}$ in Fig. \ref{fig:pr12}.
	One notes that both $M_{ss}$ and $M_{\sigma\sigma}$ exhibit violent oscillatory behavior in a certain period of time,
	which means that parametric resonance occurs for certain modes.
	Indeed, for a larger mode $k=10^3k_m$, the resonant amplification is highly efficient when the coupling terms
	in Eqs.~\eqref{eom1d} and \eqref{eom2d} are absent. It is the coupling term that weakens the parametric resonance, so that the second peak does not appear on smaller scales.
	In addition, we stress that isocurvature perturbations will not affect curvature perturbations on CMB scales due to its rapid decay for super-horizon modes.

\begin{figure*}
		\includegraphics[width=0.8\textwidth]{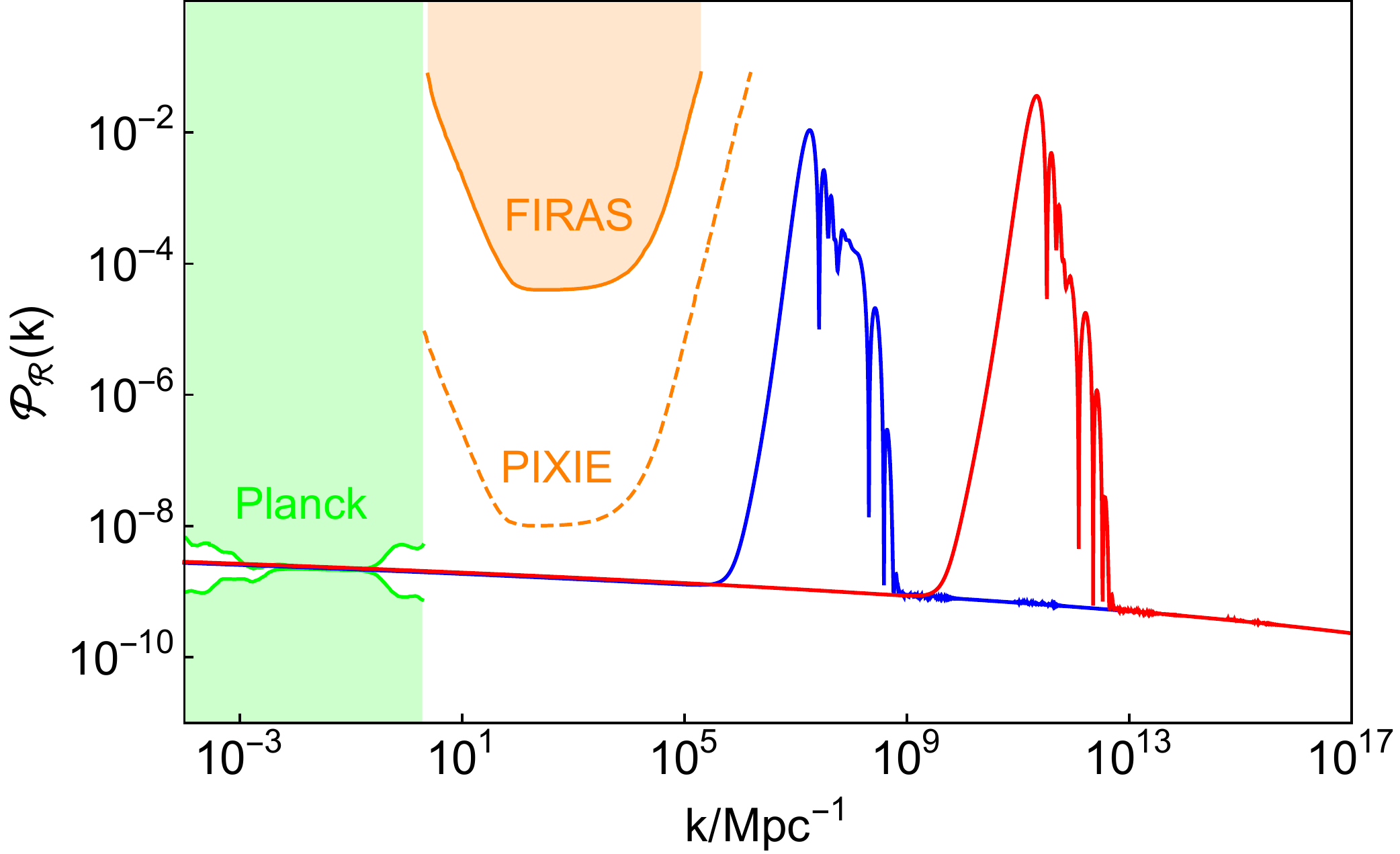}		
		\caption{Power spectra of primordial curvature perturbations.
			The blue and red lines represent the results for Set 1 and Set 2, respectively.
			The green shaded region is excluded by CMB observations~\cite{Planck:2018jri}.
			The orange shaded region shows the current upper bound on the power spectrum from measurements of $\mu$ distortion for COBE/FIRAS \cite{Mather:1993ij,Fixsen:1996nj}. The forecasted constraint for the distortion experiment PIXIE \cite{Kogut:2011xw} is shown as the orange dashed line. See Ref. \cite{Chluba:2019nxa} for the summary of constraints on the power spectrum of curvature perturbations.}
		\label{fig:pr}		
	\end{figure*}

	We present the resulting power spectra of curvature perturbations in Fig. \ref{fig:pr}, where we plot two different examples using the same potential parameters.
	\subsection{Generation of induced gravitational waves on small scales}
	\begin{figure*}
		
		\includegraphics[width=0.8\textwidth]{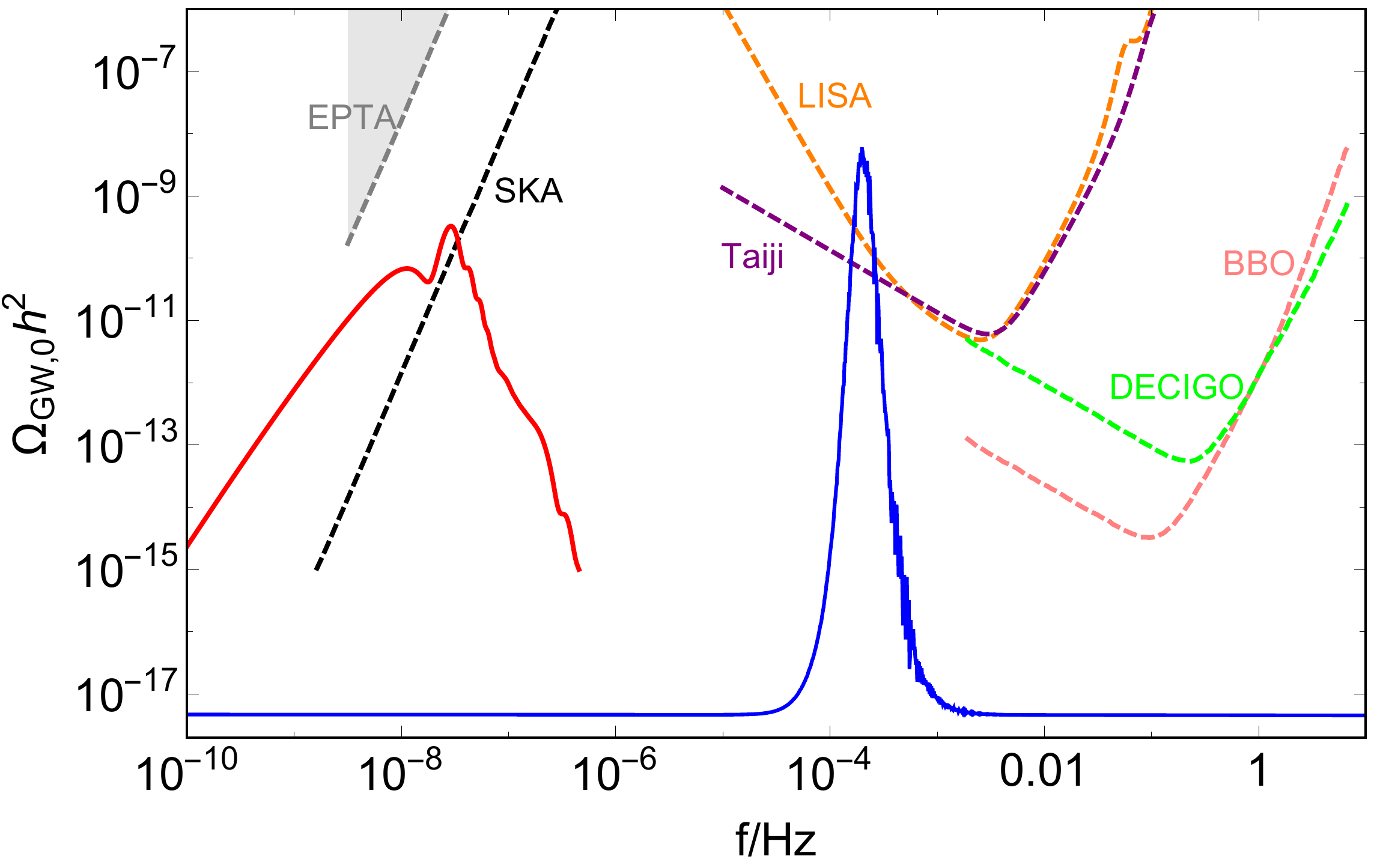}
		\caption{Present energy spectra for induced GWs and primordial GWs for Set 1, respectively.
			The red solid line represents induced GWs in the radiation-dominated era and the blue one represents primordial GWs generated during inflation.
			Other dashed lines are the expected sensitivity curve of the future GW projects summarized in \cite{Moore:2014lga}. The shaded regions represent the present existing constraints on GWs \cite{Kohri:2018awv,Lentati:2015qwp}.}
		\label{fig:omega}
	\end{figure*}	
	We now investigate GWs sourced by second-order scalar perturbations, named scalar-induced GWs. For the reasons mentioned above, we omit the detailed derivation and directly show the formula for energy spectrum of GWs, which is given by~\cite{Kohri:2018awv}
	\begin{align}\label{IGW}
		\Omega_{\rm{GW}}&(\eta,k) = \frac{1}{12} \int^\infty_0 dv \int^{|1+v|}_{|1-v|}du \left( \frac{4v^2-(1+v^2-u^2)^2}{4uv}\right)^2\mathcal{P}_\mathcal{R}(ku)\mathcal{P}_\mathcal{R}(kv)\nonumber\\
		&\left( \frac{3}{4u^3v^3}\right)^2 (u^2+v^2-3)^2\nonumber\\
		&\left\{\left[-4uv+(u^2+v^2-3) \ln\left| \frac{3-(u+v)^2}{3-(u-v)^2}\right| \right]^2  + \pi^2(u^2+v^2-3)^2\Theta(v+u-\sqrt{3})\right\}.
	\end{align}
	Taking the thermal history of the Universe into consideration, one can get the energy spectrum of GWs at present,
	\begin{equation}
		\Omega_{\rm GW,0}(k)= \Omega_{\gamma,0} \left(\frac{g_{\star,0}}{g_{\star,\rm eq}}\right)^{1/3}  \Omega_{\rm GW}(\eta_{\rm eq},k),
	\end{equation}
	where $\Omega_{\gamma,0}$ is the density parameter of radiation today, $g_{\star,0}$ and $g_{\star,\rm eq}$ are the effective numbers of relativistic degrees of freedom at the present time and at the time $\eta_{\rm eq}$ of the radiation-matter equality, respectively.

	Fig.~\ref{fig:omega} shows the current energy spectra for induced GWs and primordial GWs with parity violation for Set 1.
	We can see that the peak of the induced GWs energy spectrum locates in the sensitive region of SKA.
	It is interesting to find that the peak frequency of induced GWs is much smaller than that of primordial GWs.
	The reason for this phenomenon is almost obvious. We can estimate $k_{\rm {peak,IGW}} \simeq a(t_{\rm{osc}})H(t_{\rm {osc}})$, $k_{\rm {peak,PGW}} \simeq m_{\chi}a(t_{\rm{osc}})/2$, where $t_{osc}$ denotes the time when $\chi$ starts oscillation.
	Given that slow-roll condition $H^2 \simeq V_0/3M_{\rm p}^2$, we obtain $k_{\rm {peak,PGW}}/k_{\rm {peak,IGW}}\simeq m_\chi M_{\rm p}/\sqrt{V_0}$, which is approximately equal to $2\times 10^3$ in our parameter setting.
	Therefore, the relation of peak frequency between these two SGWBs is only dependent on the potential ratio ($\equiv m_\chi M_{\rm p}/\sqrt{V_0}$). Varying the ratio, the total SGWB spans the sensitive frequency bands of various current of future observation plans. For example, if primordial GWs with parity violation locate in the sensitive region of BBO or DECIGO, induced GWs fall in the LISA-Taiji frequency range, and so on.
	In a word, besides chirality from the Chern-Simons term, the joint measurements of these two SGWBs provide another channel to test our model.		
	
	\section{Conclusion}
	\label{V}
	Searching for signs beyond GR is an exciting topic. dCS gravity is certainly a plausible and self-consistent alternative.
	In this paper, we have investigated  GWs generated in the two-field model with the dCS coupling.
	A novelty is that we consider a very heavy field coupled with the Chern-Simons term, which leads to an increase in the third derivative of this field with respect to the cosmic time, thus, triggering the resonant amplification of primordial GWs on small scales during inflation.
	On the other hand, the heavy field forces us to consider the effects of isocurvature perturbations on curvature perturbations. It is shown that isocurvature perturbations exhibit a tachyonic instability within a short period of time, which acts as a source inducing the amplification of curvature perturbations.  As a result, there are two SGWBs located at different frequency bands in our model.
	One is amplified primordial GWs with parity violation during inflation and the other is  induced GWs sourced by curvature perturbations in the radiation-dominated era.
	The detection of these two SGWBs provides an exciting possibility to test inflationary models and gravitational theories.
	
	\begin{acknowledgments}
		We thank Wang-Wei Yu for useful discussions.
		This work is supported in part by the National Key Research and Development Program of China Grant No. 2020YFC2201501,
		in part by the National Natural Science Foundation of China under Grant No. 12075297 and No. 12235019.
	\end{acknowledgments}

	\bibliographystyle{apsrev4-1}
	\bibliography{dCSGW}
	
\end{document}